\documentclass[prx,preprint]{revtex4}
\usepackage{graphicx}
\usepackage{dcolumn}
\usepackage{bm}
\usepackage{color}
\usepackage{soul}
\usepackage{url}
\usepackage{xspace}

\newcommand{\nc}{\newcommand}
\nc{\I}{$I$}
\nc{\II}{$II$}
\nc{\III}{$III$}
\nc{\nn}{\nonumber}

\def\etal{{\em et al. }}

\nc{\XYZ }{\bf} %
\nc{\ABC }{\st}

\begin{document}
\newcommand{\av}[1]{\langle #1 \rangle}

\title{Eppur si riscalda - and yet, it (just) heats up: \\ 
Further Comments on ``Quantifying hot carrier and thermal contributions in plasmonic photocatalysis''}

\author{Yonatan Sivan$^{1,2,^\dagger}$, Joshua Baraban$^{3,^\dagger}$, Yonatan Dubi$^{2,3,^\dagger}$
\\
\normalsize{$^{1}$School of Electrical and Computer Engineering, Ben-Gurion University of the Negev, Israel}\\
\normalsize{$^{2}$ Ilse Katz Center for Nanoscale Science and Technology, Ben-Gurion University of the Negev, Israel}\\
\normalsize{$^{3}$Department of Chemistry, Ben-Gurion University of the Negev, Israel }\\
\normalsize{$^\dagger$ equal contribution}
}

\date{\today}

\begin{abstract}


Our Comment~\cite{anti-Halas-comment} (as well as Ref.~\cite{Y2-eppur-si-riscalda}, and the supporting theoretical studies~\cite{Dubi-Sivan,Dubi-Sivan-Faraday}) on recent attempts to distinguish thermal and non-thermal (``hot carrier'') contributions to plasmon-assisted photocatalysis~\cite{Halas_Science_2018} initiated a re-evaluation process of previous literature on the topic within the nano-plasmonics and chemistry communities. The Response of Zhou \etal~\cite{Halas_Science_2018-response-to-comment} attempts to defend the claims of the original paper~\cite{Halas_Science_2018}.

In this manuscript, we show that the Response \cite{Halas_Science_2018-response-to-comment} presents additional data that further validates our central criticism: inaccurately measured temperatures (that are lower than the actual temperature of the catalyst) led Zhou \etal to incorrectly claim conclusive evidence of non-thermal effects. We identify flaws in the experimental setup (e.g.~the use of the default settings for the thermal camera and incorrect positioning of the thermometer) that may have led Zhou \etal to make such claims. We further show that the Response contains several factual errors and does not address the technical problems we identified with the data acquisition in~\cite{Halas_Science_2018}. We demonstrate that both the Response~\cite{Halas_Science_2018-response-to-comment} and the original paper~\cite{Halas_Science_2018} contain additional faults, for example, in the power determination and in the normalization of the rate to the catalyst volume, and exhibit misconceptions regarding the thermo-optic response of metal nanostructures.  The burden of proof required by the proposal of a novel physical mechanism has simply not been met, especially when the existing data can be modeled exquisitely by conventional theory.

\end{abstract}

\maketitle
\tableofcontents

\section{Introduction}
\begin{figure}[h]
    \centering
    \includegraphics[width=12.5truecm]{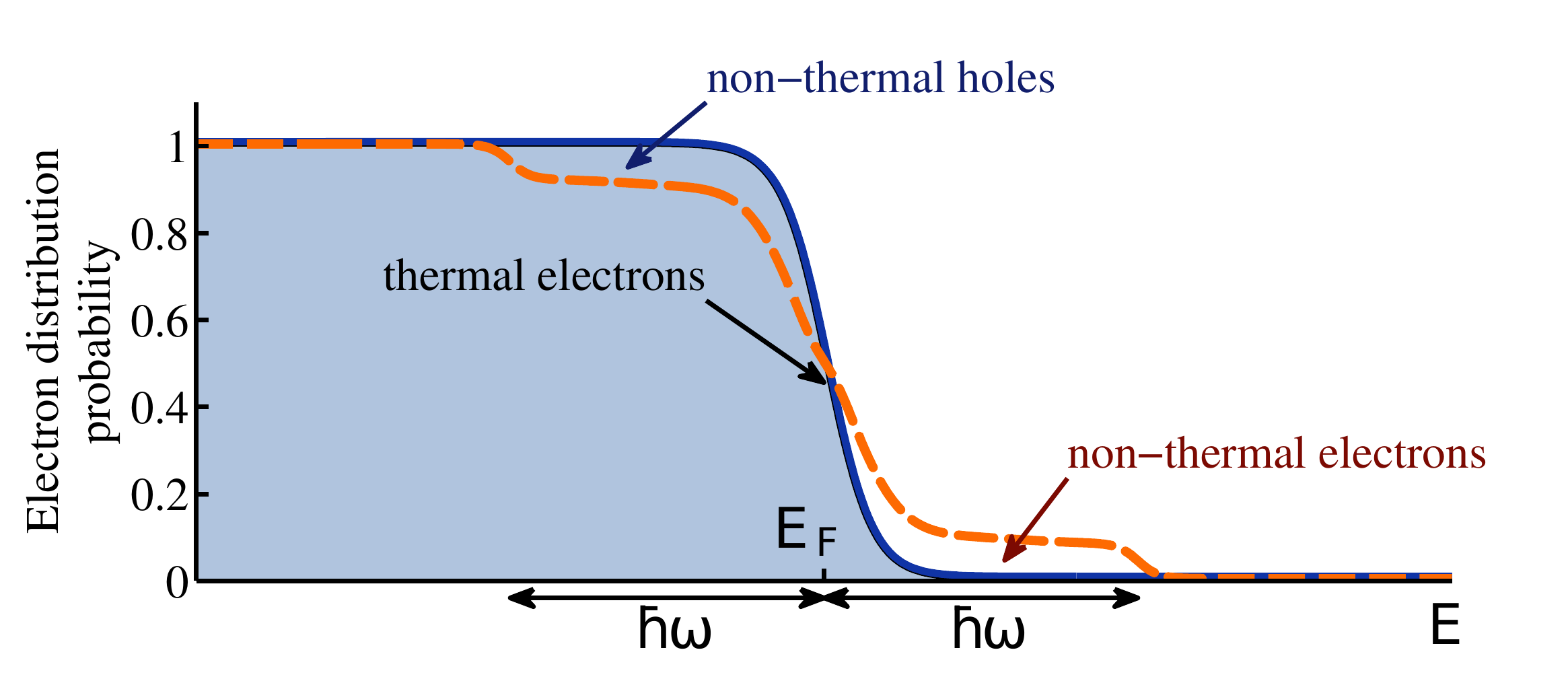}
    \caption{{\bf Schematic illustration of the electron distribution in a metal illuminated by continuous wave (CW) radiation.} The blue solid line represents the equilibrium electron distribution in the absence of illumination. The orange dashed line represents the electron distribution under illumination. It consists of thermal electrons near the Fermi energy that obey the Fermi-Dirac statistics, and non-thermal (the so-called ``hot'') electrons in two $\hbar \omega$-wide shoulders far from the Fermi energy, which are not part of the Fermi-Dirac distribution. }
    \label{fig:hot_elec_sch}
\end{figure}

Experimental demonstrations of faster chemistry in the presence of illuminated metal nanoparticles have sparked a great deal of interest among researchers in the fields of nano-plasmonics, nanophotonics, and chemistry~\cite{plasmonic-chemistry-Baffou,plasmonic_photocatalysis_Clavero,hot_es_review_2015_Moskovits,hot_es_review_2015,Valentine_hot_e_review}. Early works associated these effects with non-thermal carriers (having energies high above the Fermi energy, see Fig.~\ref{fig:hot_elec_sch}) generated upon photon absorption in the metal. However, the role of regular heating in these systems (associated with electrons with low energies with respect to the Fermi energy, also shown in Fig.~\ref{fig:hot_elec_sch}), an effect that is undesirable due to the resulting lack of selectivity and high practical costs, was not fully elucidated.

In~\cite{Halas_Science_2018}, Zhou {\em et al.} described an experiment that aimed to provide conclusive evidence for non-thermal effects. However, a {\em brief} Technical Comment~\cite{anti-Halas-comment} we published shortly after the publication of the original work identified technical and methodological issues in~\cite{Halas_Science_2018}. Specifically, we showed that incorrect measurements of the temperature likely led to an underestimation of the catalyst temperature, and that the catalytic enhancement of the reaction rates can be simply and directly attributed to illumination-induced heating using the well-known Arrhenius Law. Similar problems in additional related papers, as well as a more comprehensive discussion that includes an alternative interpretation of the experimental data, were presented in~\cite{Y2-eppur-si-riscalda}.

In their Response~\cite{Halas_Science_2018-response-to-comment} to our Comment, Zhou \etal defend their original paper, among other things, by providing additional results not reported previously.
First and foremost, we note that the newly presented data in the Response in fact {\em supports} our criticism. Specifically, the Response reports a roughly 5\% discrepancy between temperature readings from their thermal camera and from a thermocouple, whereas in \cite{Halas_Science_2018} they claim that they were identical. Setting aside several flaws in this attempted control experiment, this $\sim 5\%$ error in the temperature measurement alone adds a factor of $\sim 5$ to the thermal reaction rate, due to the exponential dependence of the reaction rate on the temperature (see Section~\ref{sec:conceptual}); this error is certainly an underestimate, and as discussed below, the remainder of the originally claimed photocatalysis enhancement of 30$\times$ could easily be erased by any of a plethora of factors.

Unfortunately, the main effort of Zhou {\em et al.}'s response consists of {\em ad hoc} attacks on our criticism, and a careful reading of their Response \cite{Halas_Science_2018-response-to-comment} reveals that it is filled with factual and scientific errors.  It effectively ignores our earlier criticism and exhibits a rather naive understanding of the temperature distribution in their sample, as well as the significant differences that exist between the photocatalysis experiment and the thermocatalysis control experiment. The response also reflects an unrealistic view of the capabilities of the experimental setup to measure the temperature and resolve its spatial variations.
We are therefore placed in the unpleasant position of being forced to explain these errors in much greater detail than in~\cite{anti-Halas-comment}, now also supported by visual evidence brought from~\cite{Halas_Science_2018} \&~\cite{Halas_Science_2018-response-to-comment}. Furthermore, so far we have refrained from noting other critical experimental flaws and technical faults beyond what we identified in~\cite{Halas_Science_2018}; these are now discussed at length. All this is done now in order to avoid these issues in the future.

In the final analysis, especially in light of the far simpler and physically well-founded explanation we proposed in~\cite{anti-Halas-comment} that exactly reproduces the experiments of~\cite{Halas_Science_2018} and others, the burden of proof rests squarely on Zhou \etal to substantiate their proposed novel phenomenon of an intensity-dependent activation energy and the dominance of non-thermal effects.

This paper is thus organized as follows. The first section is a detailed response to Zhou {\em et al.}'s response to our comment. This section may read like a review rebuttal, but it is not. Rather, it is a clarification of the errors made in \cite{Halas_Science_2018} and compounded in~\cite{Halas_Science_2018-response-to-comment}, roughly in the order presented in~\cite{Halas_Science_2018-response-to-comment}. The reader is encouraged to first read our Comment,  Ref.~\cite{anti-Halas-comment}, and then Ref.~\cite{Halas_Science_2018-response-to-comment} and Section~\ref{sec:res_to_res} paragraph by paragraph.

In the second part of this manuscript, we outline several additional problems in the data acquisition and processing of~\cite{Halas_Science_2018} (not mentioned before), which severely call into question the data itself. In the last part, we re-iterate the potentially most important message of our own set of studies on plasmon-assisted photocatalysis, namely, the severe limitation of the methodology adopted in~\cite{Halas_Science_2018} (and many others, see~\cite{Y2-eppur-si-riscalda}) to distinguish thermal {\em vs.}~non-thermal effects -- any small difference between the temperature profiles of the photocatalysis experiment and its thermocatalysis control is bound to be erroneously interpreted as a non-thermal effect. In other words, the methodology adopted so far can allow one to detect non-thermal effects only if they are far stronger than the reaction rate uncertainty associated with the temperature inaccuracy of the thermocatalysis control experiments. 



\section{Response to Response of Zhou {\em et al.}~\cite{Halas_Science_2018-response-to-comment}}\label{sec:res_to_res}

\subsection{Incorrect value for the emissivity}\label{sub:emissivity}
The first item in the Response is the chosen emissivity value. The authors of Ref.~\cite{Halas_Science_2018} used the {\sl default setting} of their thermal camera, an emissivity of 0.95 (almost an ideal black body; compare the FLIR A615 camera manual~\footnote{http://www.cctvcentersl.es/upload/Manuales/A3xxx\_A6xxx\_manual\_eng.pdf, page 85} to the settings seen on the right-hand-side of the camera image, Fig. S11 of the original work~\cite{anti-Halas-comment}, Fig.~\ref{fig:S11} below) rather than adjusting the settings to the actual experimental conditions! Indeed, their sample seems to have a much lower emissivity. Specifically, we estimated $0.02-0.2$ based on composition and structure of their sample as described in the SI of~\cite{Halas_Science_2018}. {\em Unlike what is claimed in the Response~\cite{Halas_Science_2018-response-to-comment}}, the values we used in our estimate of the emissivity of the sample {\em were not} taken for polished surfaces, but rather for small grains, see for example the MgO content values (see Fig. 18 of Ref. 5 of~\cite{anti-Halas-comment}).

To answer this, Zhou \etal state that in porous beds of powdered materials the emissivity is much larger, and refer to an example of samples consisting of Ni nanoparticles on a Al$_2$O$_3$ substrate. What Zhou \etal fail to point out is that Al$_2$O$_3$ comprises just $\sim$20 percent of their sample (see page 1 of their SI). Thus, conducting a weighted sum of the emissivities (taking the extreme limit of unity emissivity for Al$_2$O$_3$) gives a contribution of $\sim 0.2$, which is the upper limit we provided in our Comment~\cite{anti-Halas-comment}. But, this is also probably an overestimate -- the user manual for their thermal camera (page 85) 
clearly states the emissivity of powdered alumina as 0.16-0.46. To avoid this issue, the authors of Ref.~\cite{Halas_Science_2018} should have measured the emissivity (for each different sample) 
and set the thermal camera parameter to that value. 
No such attempt was made in the original paper~\cite{Halas_Science_2018}, nor in the Response~\cite{Halas_Science_2018-response-to-comment}, even though the authors did change the value for the external optics transmission factor.






As explained in our Comment, the exceedingly high emissivity setting in~\cite{Halas_Science_2018} means that the temperature readings of the thermal imaging camera underestimate the actual temperature of the sample. As shown in Section~\ref{thermocouple} below, this claim is well correlated with the incorrect choice of the thermocouple position in the newly reported benchmarking experiments (see Sec.~\ref{thermocouple} below).

\begin{figure}[ht]
\centering{\includegraphics[width=0.5325\textwidth]{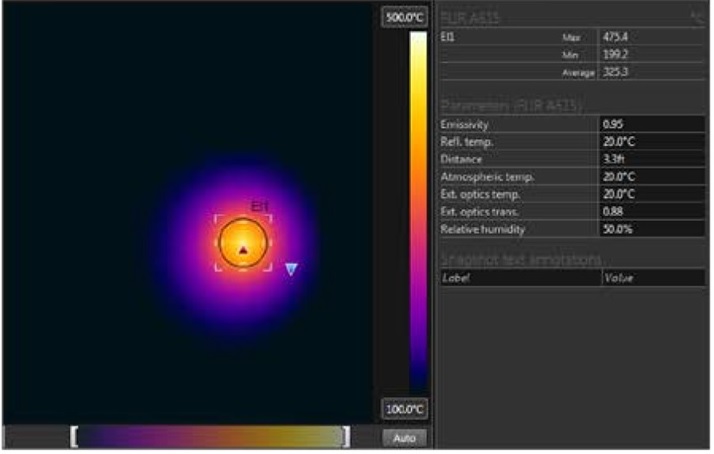}\includegraphics[width=0.485\textwidth]{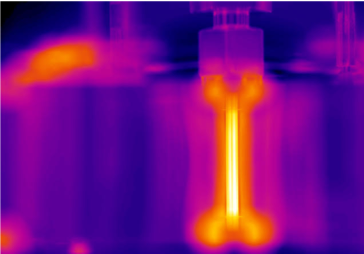}}
\caption{\label{fig:S11} ({\sl This image is best viewed as a high resolution color image}) \\ Left:~ Fig. S11 of~\cite{Halas_Science_2018}. The image is clearly blurred (out-of-focus) and the camera settings are clearly left at their default values (Distance $ = 3.3$ ft and Emissivity $= 0.95$). As explained in the text, these are very unlikely to be suitable for the experimental conditions. Right:~ As a comparison, we show an image of an object of the same size as the pellet ($2$ mm diameter tube, seen yellow-white in the center-right of the image) taken by the {\em same} camera model (equipped with a 100 $\mu m$ closeup lens, to boot). Improved image sharpness is apparent with correct focusing, but note the difficulty of obtaining sufficient resolution (and therefore accurate temperature measurements) even with the magnifying lens. Other hot(ter) objects that are out of focus seem to be cooler than their actual temperature.
} \label{Halas_SI_S11}\end{figure}

\subsection{Incorrect focussing and choice of camera-sample distance}\label{subsec:improper_focussing}
Further consideration of Fig. S11 of the original work~\cite{anti-Halas-comment} (Fig.~\ref{fig:S11}) gives rise to additional concerns. Comparison of the FLIR A615 camera manual~\footnote{https://www.flir.eu/products/a615/} to the settings seen on the right-hand-side of the camera image reveals that the (rather blurry) images were taken with the camera-to-sample distance set to the unlikely large distance of 3.3 feet, which is, again, the {\em default} setting of the camera software. 
As a comparison, Fig.~\ref{fig:S11} shows also an image of an object of the same size as the pellet ($2$ mm) taken by the {\em same} camera model. The far better achievable resolution is clearly seen. Simply put: their image is out of focus, which inevitably leads to an underestimation of temperature. The Response~\cite{Halas_Science_2018-response-to-comment} does not refer to the unlikely choices of settings for both the emissivity and camera-sample distance.



\subsection{Improper thermocouple positioning}\label{thermocouple}
\begin{figure}[ht]
\centering
\includegraphics[width=0.5\textwidth]{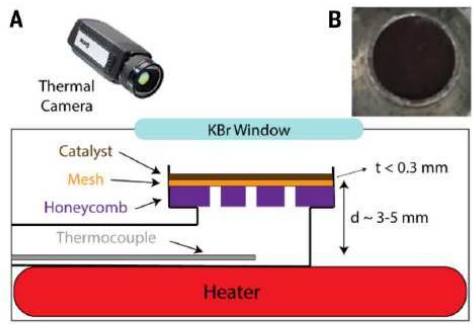}
\caption{(Color online) Fig. 1 of~\cite{Halas_Science_2018-response-to-comment}. The distance between the thermocouple and the sample is $3-5$ mm. The distance from the camera to the sample is not specified. \label{fig:config}
}\end{figure}

To validate the readings of the thermal camera, Zhou \etal compare them to those of a thermocouple, placed well below their catalytic pellet (specifically, $3-5$ mm away), see Fig.~\ref{fig:config} (a copy of a new plot that appears in the Response~\cite{Halas_Science_2018-response-to-comment}). Quite intuitively, the temperature at that position differs from the catalyst temperature: for external heating (from below), it records a temperature which is higher than the catalyst temperature, whereas for optical heating (from above), it records a lower temperature. Indeed, as we showed in~\cite{Y2-eppur-si-riscalda}, and as demonstrated experimentally by the Liu \& Everitt teams~\cite{Liu_thermal_vs_nonthermal,Liu-Everitt-Nano-Letters-2019}, even if the thermocouple is placed right at the bottom surface of the catalytic pellet in the photocatalysis experiments, it measures lower temperature compared to the top surface temperature (where the photon absorption takes place, hence, where the heat is generated)
, all the more so if the thermocouple is placed some $3$ mm below the pellet, as in~\cite{Halas_Science_2018}.

As shown in~\cite{Y2-eppur-si-riscalda}, the lower temperatures arising from the improper positioning of the thermocouple are likely to be the origin of the incorrect claims on dominance of non-thermal effects in~\cite{Halas_dissociation_H2_TiO2,Halas_H2_dissociation_SiO2}. In the current context, the fact that the temperature readings of the thermocouple and thermal camera are roughly the same only {\em strengthens} our claim that these readings refer to a temperature which is lower than what is felt by the illuminated nanoparticles (hence, of the reactants) and provide further support of our criticism on the data acquisition in~\cite{Halas_Science_2018}. Put simply, Zhou \etal compare two temperatures which {\em both} underestimate the true temperature of the catalytic pellet. The accuracy of the temperature readings is likely to be worse under illumination, where the energy is deposited in the nanoparticles on the upper surface of the pellet, leading to more pronounced temperature gradients. The authors of the Response~\cite{Halas_Science_2018-response-to-comment} did not report such a comparison of two thermometry methods under photocatalytic conditions.

What the authors {\em do} report is a $\sim 5\%$ discrepancy between the temperature readings from their thermal camera and from a thermocouple. We point that this $\sim 5\%$ error in the temperature measurement alone adds a factor of 5 to the thermal reaction rate, due to the exponential dependence of the reaction rate on the temperature (see Section~\ref{sec:conceptual}). Such a factor cannot be dismissed, especially in light of the possibility that the observed enhancement factor is largely artificial, see Sec.~\ref{subsec:normalization} below.

\subsection{More on Temperature measurements}\label{subsec:T_nonuniformities}

\subsubsection{Temporal non-uniformity}
Next, Zhou \etal discuss the uniformity of the temperature in their sample. First, considering the temperature uniformity in time (i.e., the fact that the temperature is eventually nearly constant over time even though the illumination has a pulsed nature), they refer to their Fig. S12E in the SI. This is a {\sl schematic figure}, not a measurement nor a calculation. Nevertheless, the emphasis of this issue is misleading since never in our Comment did we argue differently. In fact, we had performed the corresponding calculation and indeed found a fairly temporally-uniform temperature, see~\cite[Fig.~S4(b)]{Y2-eppur-si-riscalda}.

\subsubsection{Surface non-uniformity and spatial resolution}
After that, Zhou \etal discuss spatial variations of the temperature and point out that the values indicated in their plots is the highest surface temperature. Here they are referring to temperature variations along the surface plane, which we never mention in our Comment, so this is, again, not particularly relevant. For the sake of completeness, we note that we treated this issue at length in~\cite{Y2-eppur-si-riscalda}, and did find significant in-plane non-uniformity. Such non-uniformities might prove to be important in future studies.

However, this topic is intimately related to the crucial issue of spatial resolution, and demonstrates the unreliability of the claims made in~\cite{Halas_Science_2018}. Insufficient spatial resolution or incorrect focusing of the thermal camera (see Section~\ref{subsec:improper_focussing}) necessarily leads to temperature readings lower than the actual temperature. Zhou \etal claim a resolution of $\sim100$ $\mu m$; this contradicts the manufacturer's specifications  \footnote{https://flir.custhelp.com/app/utils/fl\_fovCalc/pn/55001-0102/ret\_url/\%252Fapp\%252Ffl\_download\_datasheets\%252Fid\%252F20}, where the resolution is given as $690$ $\mu m$ at the working distance of 3.3 ft.~indicated in Fig.~S12 of Ref.~\cite{Halas_Science_2018} (our Fig.~\ref{fig:S11}). Worse, a resolution of $\sim 100$ $\mu m$ with this camera is impossible at any working distance, according to the manufacturer. Using the correct $\sim 0.69$ mm value, we can see that a properly focussed image would achieve less than 3 pixels across the 2 mm diameter pellet, which is not adequate for accurate temperature measurement, again, according to the recommendations of the manufacturer.



\subsubsection{Depth non-uniformities}\label{subsec:depth_T_nonuniformities}
\begin{figure}[ht]
\centering
\includegraphics[width=0.7\textwidth]{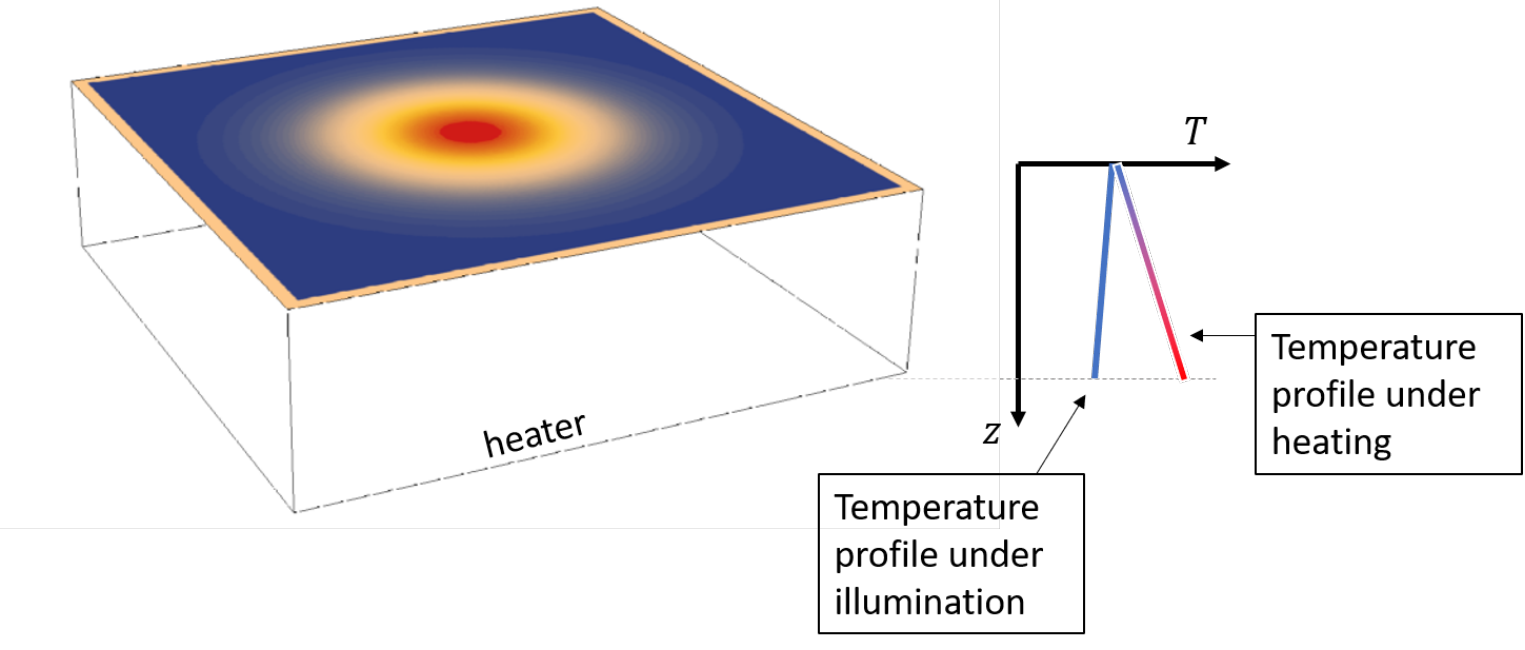}
\caption{(Color online) A schematic illustration of the temperature profile in the photocataltysis and thermocatalysis experiments. The vertical gradients are opposite in the two experiments, such that the thermocatalysis control experiment does not mimic correctly the conditions of the photocatalysis experiment.  \label{fig:T_profile}}
\end{figure}

Nevertheless, even if one ignores all the above, a less obvious yet crucial problem arises from temperature gradients along the depth of the pellet. The thermal camera gives no information about this dimension, and so we are left to rely on calculations and common sense. Our detailed temperature calculations in~\cite{Y2-eppur-si-riscalda} show that there are gradients of several {\em hundreds} of degrees across such a distance. Such gradients were also shown to exist experimentally by the Liu team~\cite{Liu_thermal_vs_nonthermal,Liu-Everitt-Nano-Letters-2019}. While it is true that, given only surface temperature information, thermocatalysis control experiments should be carried out at the highest measured temperature in order not to underestimate thermocatalytic effects, this is still inadequate in case the gradients have opposite signs, since the (top) surface temperature is lower than the bottom surface temperature in the thermocatalysis case, see Fig.~\ref{fig:T_profile}; this is exactly the case in~\cite{Halas_Science_2018}, so that clearly the control thermocatakysis experiment does not overestimate the thermal contribution in the photocatalysis case. Moreover, the Liu/Everitt teams~\cite{Liu_thermal_vs_nonthermal,Liu-Everitt-Nano-Letters-2019} discussed additional consequences in the reaction rate caused by these opposing gradients. Thus, an effective control experiment must ensure the temperature profiles in the photocatalysis and thermocatalysis experiments are exactly the same (see also discussion in Section~\ref{sec:conceptual}).

\subsection{Intensity-dependence of temperature}
Zhou \etal continue their Response by criticizing the intensity-dependence of the temperature in our model, specifically, that our model {\em a priori} assumes an incorrect linear relation between the temperature and the illumination intensity. Here, they commit a few factual and physical {\em faux pas}.

First, {\em unlike what is said in the Response}, in our model we never assume {\em a priori} that the temperature is linear in intensity. Quite the opposite -- we start by assuming a general form,
\begin{equation}
T(I_{inc}) = T_0 + a I_{inc} + b I_{inc}^2.
\end{equation}
Clearly, $T_0$ is the temperature of the sample in the dark. What we find from fitting this expression (placed in the Arrhenius formula) to {\sl their} data (specifically, Figs. 2(a)-(b) in~\cite{Halas_Science_2018}), is that the nonlinear term is vanishingly small. Therefore, the linear dependence in Fig.~1 of our Comment~\cite{anti-Halas-comment} comes out of the data rather than being pre-assumed. 

In this context, the linear model works perfectly well for most of the data presented in~\cite{Halas_Science_2018} (as well as for all the data of several other papers we criticize in~\cite{Y2-eppur-si-riscalda}, see discussion in~\cite[p. 270-271]{thm_hot_e_faraday_discuss_2019}). However, like any other physical system, the range of validity of the linear response is finite. Zhou \etal quote Refs.~10-11 in their comment as stating that a linear dependence of the temperature on the illumination is only applicable for temperature rises smaller than about a 100K. This is never stated in neither of these references; these papers discuss only the linear relation (which {\em is} criticized by Zhou \etal). Instead, these claims in fact originate from papers of the Sivan group~\cite{Sivan-Chu-high-T-nl-plasmonics,Gurwich-Sivan-CW-nlty-metal_NP}, see below, which Zhou \etal fail to quote. Either way, these claims were raised in~\cite{Sivan-Chu-high-T-nl-plasmonics,Gurwich-Sivan-CW-nlty-metal_NP} in the quite different context of a single illuminated NP configuration, where temperature-induced changes to the metal permittivity are dominant. As shown below, in the current context of a mm-scale composite that contains a very large number of sparsely dispersed NPs, the nonlinear thermo-optic response originates from the host (as it occupies the vast majority of the sample volume), see also~\cite{Donner_thermal_lensing}. Since the permittivity of dielectric materials is far less sensitive to heat, the nonlinear thermo-optic response manifests itself at much higher temperatures - several hundreds of degrees. This is in correlation with the observation of Zhou \etal in Fig.~1(d) of~\cite{Halas_Science_2018}.


An additional error made by Zhou \etal. is the formula they suggested for analyzing the dependence of the temperature on illumination, namely,
\begin{equation}
\alpha I = h(T) (T-T_0) + A(T) (T^4 - T_0^4).
\end{equation}
This equation is essentially energy conservation - it equates absorption of photons (left-hand-side) to radiative (quartic term) and non-radiative (linear term; heat conductance) heat loss (right-hand-side). What Zhou \etal get wrong is that the latter is extremely small. In particular, a quick evaluation (see Appendix~\ref{AppA}) reveals that heat radiation is at least about $10^4-10^5$ times smaller than the power that is lost via thermal conductance - because their nanoparticles are not isolated. 
Put simply - the power that is lost due to radiative heat losses is only a tiny fraction of the power that is lost through direct contact between the nanoparticles and the substrate, so that the nonlinearity has nothing to do with radiative heat losses.

Instead, the nonlinearity has two main sources, both much stronger than radiative losses. First, the absorption coefficient $\alpha$ depends on the temperature {\em via} the temperature dependence of the metal permittivity, an effect studied in countless papers, see e.g.,~\cite{Langbein_PRB_2012,Stoll_review} for the ultrafast temperature dependence of the metal permittivity; many other papers, including various ellipsometry papers, studied the corresponding steady-state temperature dependence, see e.g.,~\cite{Shalaev_ellipsometry_gold,Shalaev_ellipsometry_silver,Sivan-Chu-high-T-nl-plasmonics,Gurwich-Sivan-CW-nlty-metal_NP,PT_Shen_ellipsometry_gold}, to name just a few. Second, the heat transfer coefficient $h(T)$ also depends on the temperature via e.g., the thermal conductivity, Kapitza resistance etc. (see e.g.,~\cite{Sivan-Chu-high-T-nl-plasmonics,Gurwich-Sivan-CW-nlty-metal_NP}). The exact quantification of these nonlinear thermo-optic effects is a topic which has occupied the Sivan group in the last few years (see~\cite{Sivan-Chu-high-T-nl-plasmonics,Gurwich-Sivan-CW-nlty-metal_NP,PT_Shen_ellipsometry_gold,ICFO_Sivan_metal_diffusion}); we are currently in the process of quantifying these two effects in the current context of plasmon-assisted photocatalysis (namely, a calculation of $a$ and $b$ from first principle modelling and matching them to the experimental data), and expect to publish first results soon. 

\subsection{Nanoparticle melting}
In a direct continuation of their reasoning, Zhou \etal point out that within the model we present, the temperatures would rise above the melting temperature of the copper naonparticles, thus leading to sintering which was not observed. To answer this, one needs to consider the following points. (1) Melting is an ambiguous concept for the small NPs employed in~\cite{Halas_Science_2018}; one may argue that melting occurs even under the conditions reported in~\cite{Halas_Science_2018} itself. (2) Since the nanoparticles are embedded within a porous substrate, they are separated from each other by an oxide layer and air, which may prevent sintering. (3) The authors state that no sintering was observed, but do not show data to support this claim. (4) As explained in the previous section, the main results of the original paper~\cite[Fig.~2]{Halas_Science_2018} are limited to illumination intensities $I \leq 4$W cm$^{-2}$, which, according to our fitting, lead to temperatures still below (an approximate) melting point. In particular, it is not clear why the authors do not show data points for higher intensities (except in Fig.~\cite[Fig.~1(d)]{Halas_Science_2018}). (5) The data that the authors do show that includes higher intensities (Fig.~1D and S11 in the SI of~\cite{Halas_Science_2018}) shows the onset of nonlinearity at roughly the intensity which presumably leads to heating by several hundreds of degrees, where melting might be expected. We refer the interested reader to a far more thorough discussion of this issue in~\cite[p. 270 and on]{thm_hot_e_faraday_discuss_2019}. Overall, all the above points out that even if melting occurred, it is not likely to have modified the thermal/optical/chemical performance of the pellet.

\subsection{Intensity-dependent activation energy}
In the final part of their response, Zhou \etal state that ``the assumption of a light-independent $E_a$ is not physical, because hot carriers modify adsorbate coverage on the catalyst surface and thus influence the apparent activation barrier, as we explained in our original paper''. It is hard to follow their reasoning here, since the only proof they provide for this statement is data that can be fitted -- to remarkable accuracy -- with a light-independent $E_a$.

Finally, Zhou \etal point that even if one assumes an intensity-dependent temperature, the evaluated activation energy $E_a$ still depends on the illumination intensity. This is simply incorrect, because if one assumed both intensity-dependent temperature (e.g., $T(I) = T_0 + a I$) as well as intensity-dependent activation energy $E_a(I)$, then the data would not be sufficient to determine both of them. Simply put, one can choose any value for $a$, from $a=0$ up to our value of $a \sim 180 $K/W cm$^{-2}$ (and higher), and obtain - from the same data - a different curve for $E_a(I)$. These $E_a(I)$ curves changes for different values of  $a$,  from the curve shown in Fig.~2 of ~\cite{Halas_Science_2018}, up to $E_a$ which is essentially intensity-independent, and all this with remarkable accuracy. 

\begin{figure}[ht]
\centering
\includegraphics[width=0.7\textwidth]{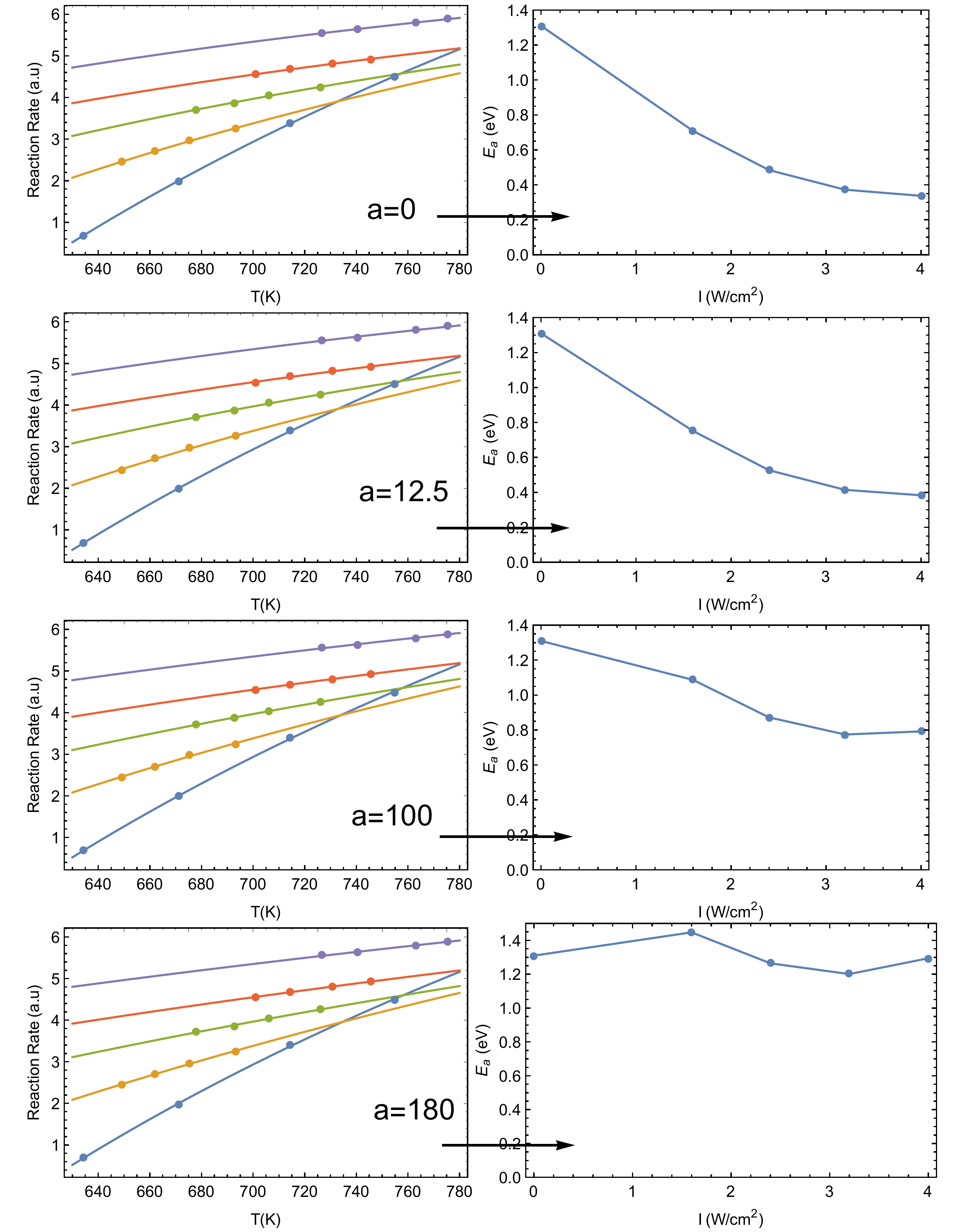}
\caption{(Color online) Left panel: reaction rate as a function (inverse) temperature, points are data from Zhou \etal \cite{Halas_Science_2018} (the data is available in Appendix \ref{AppB}). The solid lines are fits to an Arrhenius form with varying values of $a$. Right panel: the resulting activation energy as a function of intensity, going from a strongly intensity-dependent activation energy (this is what is plotted in Fig.~2C of Ref.~\cite{Halas_Science_2018}), all the way to an essentially intensity-independent activation energy for $a=180$ K/W cm$^{-2}$. \label{fig:fitting_a}
}\end{figure}

An example of this fitting procedure is shown in Fig.~\ref{fig:fitting_a}. On the left panels we plot the original data (reaction rate vs temperature) of Ref.~\cite{Halas_Science_2018} (blue points are reaction rate in the dark, and yellow, green, red and purple are for intensities $I_{inc} = 0,1.6,2.4,3.2$ and $4$ W/cm$^2$. The solid lines are fits to an Arrhenius form, $R = R_0 ~\mathrm{exp}\left(-\frac{E_a}{k_B (T_0 + a I_{inc})} \right)$, where $a$ takes different values, from $a = 0$ (as chosen by Zhou \etal) up to $a=180$K/W cm$^{-2}$ (left panels). The right panel shows the corresponding $E_a(I)$-curves, ranging from a strongly intensity-dependent activation energy (at $a=0$) to an essentially intensity-independent activation energy, all obtained with the same data and the same accuracy. This demonstrates that the last statement in the response of Zhou \etal is simply wrong. For the reader's convenience, we have added in Appendix~\ref{AppB} the data used to extract these fits (which was obtained by digitizing the original figures of Ref.~\cite{Halas_Science_2018}). The active reader can simply take these data, fit them to an Arrhenius form and see the remarkable agreement.

\section{Additional problems}
Up till now, we have only addressed points mentioned in the Response of Zhou \etal~\cite{Halas_Science_2018-response-to-comment} to our Comment \cite{anti-Halas-comment}. We now discuss several issues which were not treated in our original Comment, yet cast further doubt on the original claims of Zhou \etal of isolating the non-thermal effects from the thermal effects~\cite{Halas_Science_2018}.

\subsection{Uncertainties in the illumination intensity}
Here follow a few particularly concerning ambiguities regarding the illumination source used in Ref.~\cite{Halas_Science_2018}.

In the Supporting Material of \cite[pg.~5]{Halas_Science_2018}, Zhou \etal write, ``Direct illumination of the thermal camera with our light source did not cause any response to temperature, demonstrating that the illumination source has no mid-IR ($2-10\ \mu$m) photons.''  Given that (a) the long wavelength infrared camera's response range is $7.5-14\ \mu$m \footnote{https://support.flir.com/DsDownload/Assets/55001-0102-en-US.html}, and (b) the specification sheet of the light source states and shows that the output extends well above $2\ \mu$m \footnote{https://www.findlight.net/front-media/products/datasheet/WhiteLase\_SC400\_UV\_v1.pdf}, this statement is both factually incorrect and logically indefensible.  
In particular, it is conceivable that a massive amount of infrared radiation also illuminated the catalyst pellet during these experiments. According to the manufacturer's specifications, the light source outputs 8 W total power from $<$$410$ to $\approx$$2400$ nm \footnote{https://www.gophotonics.com/products/lasers/nkt-photonics/29-650-wl-sc-400-8},\footnote{https://www.nktphotonics.com/lasers-fibers/product/superk-extreme-fianium-supercontinuum-lasers},\footnote{https://www.nktphotonics.com/wp-content/uploads/sites/3/2018/12/superk-ext-fiu-spectral-power-density-v2.jpg}; Zhou \etal describe the use of a filter to reduce the power to 300 mW, but it is not clear if that value was explicitly measured, and if so, how. If the 300 mW value relied on the filter specifications, then an inspection of Edmund Optics' catalog shows that their relevant filters are not specified past 1200 nm; none of the other optical components used (KBr window, N-BK7/SF5 lens) would have blocked the substantial infrared power ($3-4\times$ the UV/Visible power) produced by the laser.

Furthermore, the output beam profile is Gaussian, meaning that its peak intensity is roughly twice the average intensity value taken by Zhou \etal in their data analysis.  This non-uniform illumination intensity would only exacerbate the problem of temperature gradients and non-uniform heating; again, that leads to more significant thermocatalysis effects in the photocatalytic experiments.

Finally, it is curious that the specified collimated output beam diameter of the laser in the visible is equal to or even smaller than (1.5 mm @ 530 nm) the 2 mm spot size to which Zhou \etal claim to have focused it with a fairly short focal length ($f=100$ mm) lens. 

\subsection{Normalization of the reaction rate}\label{subsec:normalization}

One of the most problematic aspects in the original paper~\cite{Halas_Science_2018} which was not raised in our Comment~\cite{anti-Halas-comment} is worth discussing now. The main claim in~\cite{Halas_Science_2018} relies on a single post-processing procedure - the rescaling of the volume contributing to the reaction according to the penetration depth of the electric field; this is justified by claiming that the contribution of non-thermal electrons can come only from the illuminated layer, whereas for thermocatalysis, the contribution to the reaction is supposed to come from the whole layer thickness. In particular, the authors estimate the penetration depth, and renormalize the reaction rate accordingly by a factor of $\sim 30$ according to~\cite{Halas_Science_2018} (or $20-100$ according to~\cite{Halas_Science_2018-response-to-comment}). This estimate is rather crude - the electric field decays exponentially, and the electron distribution scales with its square (which means that the decay occurs at twice as short a distance); this is different from the step-like dependence assumed by this normalization. A proper integration is called for - it will show that the exponential weight makes the regions of highest field matter much more (see e.g., the procedure described in~\cite{Liu_thermal_vs_nonthermal,Liu-Everitt-Nano-Letters-2019}). As shown in Section~\ref{sec:conceptual}, even small errors associated with crude normalization might result in a very large effect on the reaction rate.  Moreover, this approach ignores two additional complications. First, the temperature penetration is also finite (see our detailed calculations in~\cite{Y2-eppur-si-riscalda}) so that a similar rescaling should have been applied to the thermocatalysis. As discussed in Section~\ref{subsec:depth_T_nonuniformities}, the experimental set up provides no information on this aspect thus introducing further significant errors to the data. 
Second, these two penetration depths can vary significantly with the temperature due to the temperature dependence of the various thermal and optical parameters of the pellet constituents~\cite{Sivan-Chu-high-T-nl-plasmonics}. Indeed, changes of several tens of percent were observed in these quantities due to elevated temperatures (e.g., in~\cite{plasmonic-SAX-ACS_phot,japanese_size_reduction}). 

The bottom line is that the factor by which the authors claim that the photocatalysis is higher than thermocatalysis is very similar to the value by which the thermocatalysis was normalized. Thus, if this normalization procedure had not been used, essentially no difference between reactions rates under illumination or in the dark would have been observed. In that sense, the normalization must be very accurate in order to allow extracting valid conclusions. Moreover, the normalization voids the claims in the Response about overestimation of the thermocatalysis, see Section~\ref{subsec:depth_T_nonuniformities}. In that sense, our alternative (normalization-free) explanation that there is a negligible contribution of the non-thermal electrons to the reaction sounds far more likely compared with the crude and even somewhat artificial rescaling of the reaction data.



\section{Methodological limitations of~\cite{Halas_Science_2018}}\label{sec:conceptual}
Now that it is clear that the temperature (and likely, the intensity and reaction rate) readings in~\cite{Halas_Science_2018} are wrong, we should note that the unacceptable errors in the temperature measurements are only a prelude to the real problem of the original paper~\cite{Halas_Science_2018}! Indeed, since the same (incorrectly) {\em measured} temperature was used also for the control thermocatalysis experiments, one could claim (the authors of the response have not...) that a simple remedy to the temperature errors is to rescale the temperature axes in all plots to the correct values. Then, the observation of a faster reaction in the photocatalysis experiments could still be claimed as proof for the high energy, non-thermal carrier action.

However, here comes the more fundamental criticism we raised in~\cite{anti-Halas-comment} over~\cite{Halas_Science_2018}. To repeat, we claimed that any tiny difference between the temperature distribution in the thermocatalysis and photocatalysis experiments would immediately be interpreted as due to "hot" electrons. Further, even a tiny difference in the temperature will be significantly amplified due to the {\em exponential sensitivity} of the reaction rate to the temperature (see Section~\ref{sec:conceptual}). Indeed, for the reported activation energy of $\mathcal{E}_a = 1.3$eV and a temperature of $k_B T \sim 0.03$eV, it is easy to check that even a few percent changes in the temperature (like those reported in Table 1 of~\cite{Halas_Science_2018-response-to-comment}) lead to reaction changes of hundreds of percent.

One obvious reason for the difference in the temperature profiles in the photocatalysis and thermocatalysis experiments is provided in the plot of the response itself~\cite{Halas_Science_2018-response-to-comment}. It shows that the configuration in these two experiments is different - in the thermocatalysis experiment, heating is done from below while in the photocatalysis, the heat due to light absorption is deposited onto a thin layer on the top of the pellet surface. Thus, it is obvious that the temperature profile in the thermocatalysis and photocatalysis experiments cannot be identical, see Fig.~\ref{fig:T_profile}. In fact, these changes were shown explicitly in~\cite{Liu-Everitt-Nano-Letters-2019} to cause significant differences in the reaction rates. 


Remarkably, the Response~\cite{Halas_Science_2018-response-to-comment} ignores this methodological limitation. 




\section{Summary}\label{sec:summary}


We conclude that the response of Zhou {\em et al.} is superficial, and does not refer to the core of our criticism - the use of default settings of thermal camera, the failure of a distant thermo-couple to measure correct temperature, exponential sensitivity and the highly reasonable thermal calculations.


\appendix

\section{Estimation of radiative vs non-radiative heat transfer} \label{AppA}

The energy conservation equation of Zhou \etal reads
\begin{equation}
\alpha I = h(T)(T - T_0) + A(T) (T^4 - T_0^4).
\end{equation}
The first term describes the contact thermal conductance, i.e., heat transfer via vibrations of the solid at contact (to be referred to below as ``non-radiative'' heat transfer), and the second term describes the radiative heat loss, i.e., the heat transfer due to black body radiation absorption and emission (the the Stephan-Boltzmann law). Zhou {\em et al.} claim that the second term is responsible for the apparent nonlinear dependence on the temperature observed in their data.

It is very easy to make an estimate of the importance of the two terms, to see that the second term is much smaller, and hence has nothing to do with the nonlinearity. Specifically, the radiative power output per unit area is
\begin{equation}
P_{rad}/A = \sigma (T^4 - T_0^4),
\end{equation}
with $\sigma = 5.67 \cdot 10^{-8}$ W/m$^2$ K$^4$. This is an upper limit, assuming that the emissivity is 1 (although it is likely not, see Section~\ref{sub:emissivity}). The non-radiative heat transfer, which is the power per unit volume that goes from one hot body to another via the vibrations of the molecules, can be estimated by
\begin{equation}
P_{non-rad}/V = G_{ph-env} (T - T_0),
\end{equation}
where $G_{ph-env}$ is the typical thermal conductance of the host. The typical thermal conductance of the nanoparticles can be estimated as $G_{ph-env} \sim 5 \cdot 10^{14}$ W/m$^3$ K~\cite{Dubi-Sivan}. Thus, even if we take extremely small nanoparticles, with a typical dimension of 2nm, the contribution of the second term is $P_{non-rad} / A \sim (T-T_0)\times 10^6$ W/m$^2$ K. Comparing these terms at, say, $T = 700$K, gives $P_{rad}/A \sim 13000$ W/m$^2$, $P_{non-rad}/A = 4 \cdot 10^8$ W/m$^2$, about 4 orders of magnitude difference in favor of non-radiative heat transfer. In fact, in contrast to the claim in the Response, these terms become comparable only for temperatures as high as 20,000K(!) (rather than for a few hundreds of degrees).

\section{Reaction rate data reproducing Eq.~(\ref{fig:fitting_a})}\label{AppB}
For the reader's convenience, we add here the data extracted from Ref.~\cite{Halas_Science_2018}. The intrepid reader who has made it this far is encouraged to repeat the calculation we present (which is very simple), and to reproduce the remarkable fits of the data to the Arrhenius form.
\bigskip

\begin{table}[h!]
\begin{tabular}{cccccc}
$\text{I=0}$ & $\text{}$ & $\text{I=1.6 W/}\text{cm}^2$ & $\text{} $&$ \text{I=2.4 W/}\text{cm}^2$ & $\text{}$ \\
$T^{-1}$ & $\text{Reaction Rate}$ &$ T^{-1}$ &$ \text{Reaction Rate}$ & $ T^{-1}$ & $\text{Reaction Rate}$  \\
0.00132487 & 4.49512 & 0.00144277 & 3.24939 &  0.00137739 & 4.25341\\
0.00139989 & 3.39814 & 0.00148028 & 2.9705 & 0.00141597 & 4.04889\\
0.00148992 & 1.98507 &0.00151029 & 2.71019 & 0.00144384 & 3.86296\\
0.00157674 & 0.683559 &0.0015403 & 2.44989 &0.00147492 & 3.7142\\
\end{tabular}
\end{table}

\bigskip\bigskip

\begin{table}[h!]
\begin{tabular}{cccc}
$\text{I=3.2 W/}\text{cm}^2 $& $\text{} $& $\text{I=4 W/}\text{cm}^2 $& $\text{} $\\
$T^{-1}$ & $\text{Reaction Rate}$ & $T^{-1}$ & $\text{Reaction Rate} $ \\
0.00134094 & 4.92276 & 0.0012895 & 5.8896 \\
0.00136881 & 4.8112 &0.00131093 & 5.79663 \\
0.00139989 & 4.68105 & 0.00135059 & 5.62929\\
0.00142669 & 4.5509 & 0.00137631 & 5.55492 \\
\end{tabular}
\end{table}

\end{document}